\documentclass{emulateapj}

\shorttitle{Environmental impact of LBGs}
\shortauthors{TASKER AND BRYAN}

\begin{document}

\title{The environmental impact of Lyman-break galaxies}

\author{Elizabeth J. Tasker\altaffilmark{1,2}, Greg L. Bryan\altaffilmark{1}}

\altaffiltext{1}{Department of Astronomy, Columbia University, New York, NY 10027}
\altaffiltext{2}{Oxford University, Astrophysics, Keble Road, OX1 3RH, UK}

\begin{abstract}

We perform cosmological simulations of galaxies forming at $z=3$ using the hydrodynamics grid code, \emph{Enzo}. By selecting the largest galaxies in the volume to correspond to Lyman-break galaxies, we construct observational spectra of the HI flux distribution around these objects, as well as column densities of CIV and OVI throughout a refined region. We successfully reproduce the most recent observations of the mean HI flux in the close vicinity of Lyman-break galaxies but see no evidence for the proximity effect in earlier observations. While our galaxies do return metals to the IGM, their quantity and volume appears to be somewhat less than observed. We conclude that either we do not adequately resolve galactic winds, or that at least some of the intergalactic metal enrichment is by early epoch objects whose mass is smaller than our minimum resolved halo mass.

\end{abstract}


\maketitle

\section{Introduction}

The diverse structure and myriad of elements in the intergalactic medium (IGM) are witness to the strong role of stellar feedback in galaxies. What is not clear is which class of objects is primarily responsible for the enrichment.  One set of likely candidates are the early forming Lyman-break galaxies (LBGs). These galaxies are found to be highly clustered, with high star formation rates and possess strong winds with velocities up to 775\,km s$^{-1}$ \citep{Pettini2001, Pettini2002}.

Whether LBGs are the cause of the intergalactic metals is hotly contested. The issue hangs on whether the metals were produced by massive galaxies such as the LBGs, or whether they originated in smaller objects at even earlier times, such as dwarf galaxies or population III stars. There is evidence to support both sides. Results from \citet{Adelberger2003,Adelberger2005} (hereafter A03 and A05) find a cross-correlation between CIV systems and galaxies that is very similar to the autocorrelation function between LBGs. Additionally, gas that lies within 40\,kpc of LBGs contains strong CIV absorption lines and there is a similar association with the OVI systems. This is evidence that the LBGs might be the source of the metal systems. However, studies of the CIV column density between the redshifts of $1.5-5$ \citep{Songaila2001, Schaye2003} seem to show relatively little variation in the amount of carbon in the IGM over time, suggesting that the IGM metals were already in place at the highest observable redshifts, and must therefore stem from the second option of small, early epoch objects. 

One query surrounding LBGs is whether they are able to thrust the metals they produce out of their gravitational pull and into the IGM. Observational studies of the LBGs have suggested that outflows from the galaxies are strong enough to bodily displace the gas and A03 observed a possible void in the absorbing gas at distances out to 1\,h$^{-1}$Mpc from the galaxy. This \emph{proximity effect} results in an increase in the transmitted flux near the galaxy, where one would intuitively expect to see a decrease as the galaxy is approached. The most likely mechanism proposed for this was strong kinetic winds \citep{Croft2002}, which seems plausible based on the observations of winds by \citet{Pettini2002}. However, recent simulation work has shown these winds have little effect on the absorption at these distances \citep{Kollmeier2005, Bruscoli2003} and later observational work in this area (A05) has failed to confirm the increase in flux. Whether LBGs are therefore capable of dispersing metals into the IGM remains an open question. 

In this letter we perform cosmological simulations of galaxies forming at $z=3$. By constructing spectra that would be observed along multiple lines of sight from a distant quasi-stellar object, we measure the flux close to the galaxies and the column densities of HI, CIV and OVI in the IGM. By comparing these results with observations, we search for signs of a proximity effect surrounding LBGs and aim to determine the most likely cause of metal enrichment in the IGM.

\section{Computational Methods}

Our simulations were performed using the hydrodynamics adaptive mesh code, \emph{Enzo} \citep{Bryan1999, OShea2004}. We used a box with a comoving length of $20$\,h$^{-1}$Mpc and a $\Lambda$CDM cosmology with $(\Omega_\Lambda, \Omega_{\rm DM}, \Omega_{\rm b}, h, \sigma_8) = (0.7,0.26, 0.04, 0.67, 0.9)$

The simulation was evolved through to $z=3$ and the location of the halos was found using a halo finder developed by \citet{Eisenstein1998}. Initially, a low resolution run of the whole box was performed and the halo positions located. The simulation was then repeated with a section of approximately 5\,h$^{-1}$Mpc resolved to a maximum resolution of 1.8\,kpc with a dark matter particle mass of $5\times 10^6$\,M$_\odot$.  This translates into a minimum resolved halo mass of approximately $\sim 5\times 10^9$\,M$_\odot$.

The star particles formed in the simulation return thermal energy and metals back to the baryons. The energy is put back into the thermal energy of the gas over a dynamical time and is equal to $10^{-5}$ of the rest mass energy of generated stars, equivalent to one supernovae of $10^{51}$\,ergs per 55\,M$_\odot$ formed.  It is known that some of this energy is radiated away, although previous work with this code \citep{Tassis2003} has shown that it can eject gas from halos.  Metals are returned to the gas over the same time period with a yield of 0.02.  Radiative cooling is computed including metal line cooling based on the local metal content. To convert the metallicity into the ionization fractions for CIV and OVI, we used CLOUDY \citep{Ferland1998}, assuming ionization spectrums from \citet{Haardt1996} and, for the softer UV background, \citet{Haardt2001}

\begin{figure}[!t]
\centering
\includegraphics[width=8.5cm]{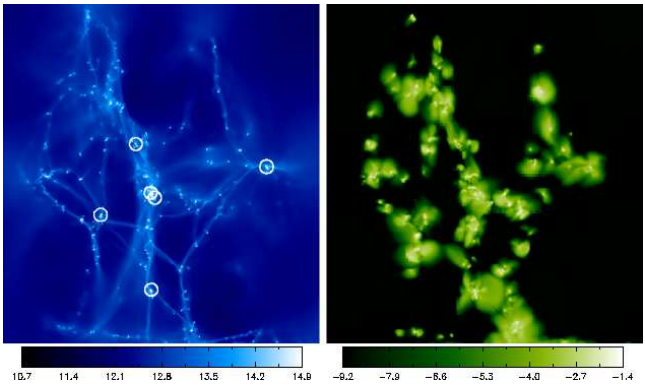}
\caption{Projections of the baryon density (left) and the fractional metallicity, $\rho_{\rm metal}/\rho_{\rm total}$. The white circles on the baryon density image show the location of the six Lyman-break galaxies. Note that the apparent close proximity of the central two galaxies is misleading, due to the image being projected down the $y$-axis. Image size is $5.31\times 4.69$\,h$^{-1}$Mpc with an integrated depth of $4.38$\,h$^{-1}$Mpc. Scales are logarithmic with baryon units M$_\odot$Mpc$^{-2}$.}
\label{fig:projections}
\end{figure}

Ideally, the LBGs in our simulation box would be chosen based on a calculation of their spectrum in the U-band, as with observations. However, for simplicity we selected the halos whose mass was the same order as the observed halo masses of LBGs, $10^{11.5\pm 0.3}$\,M$_\odot$ (A05). With this lower limit, we get a space density over the simulation box of $6.75\times 10^{-3}$\,h$^3$Mpc$^{-3}$ in good agreement with the observed value of $8\times 10^{-3}$\,h$^3$Mpc$^{-3}$ \citep{Adelberger1998}. Within our refined region (which is actually overdense with respect to the rest of the box), we get 6 halos that are above the cutoff mass.  These are circled in white in the left-hand panel of Figure \ref{fig:projections}.

\section{Results}

Figure~\ref{fig:projections} shows the baryon density (left panel) and metallicity of the refined region containing 6 halos we classify as LBGs. We can see that the LBGs are situated at the intersection of multiple gas filaments, and so mark out the densest areas of the simulation box in agreement with the observed highly biased distribution \citep{Adelberger1998}. They are also a source of metals, with each region round the LBG being surrounded by a high metallicity bubble, as shown in the right-hand panel of Figure~\ref{fig:projections}. 

\subsection{Flux properties}

\begin{figure}
\begin{center}
\includegraphics[width=7.0cm, angle=270]{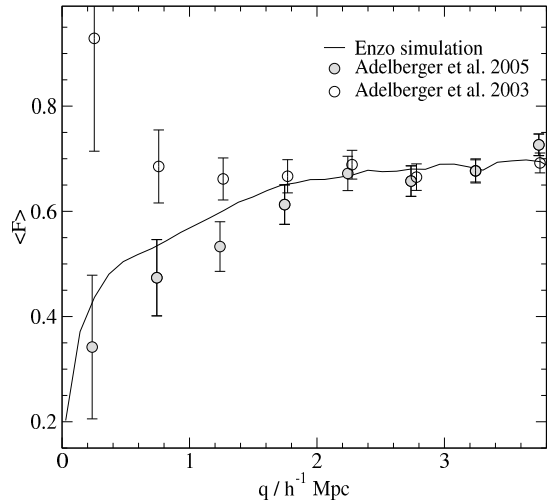}
\caption{Mean transmitted flux as a function of distance from the Lyman-break galaxies. The symbols mark out the observational results from A03 (open circles) and A05 (filled circles). Our simulation results are shown by the solid line and agree well with the most recent data set. All plots are normalised to agree with the 2005 (A05) observations at $q/h^{-1}= 3.25$\,Mpc.}
\label{fig:flux}
\end{center}
\end{figure}

To analyse the distribution of the HI near to the LBGs in our simulation, we computed the optical depth along 2000 randomly chosen lines of sight. By calculating the observed distance to each galaxy along the lines of sight, the average flux, $\left<F\right>=e^{-\tau}$, as a function of distance from the galaxy, $q$, was measured. The flux was divided and averaged into just under 50 bins and these results are presented in Figure~\ref{fig:flux}. Since the exact amplitude of the background ionizing radiation is uncertain, the results are scaled to match Adelberger's 2005 results at $q=3.25$\,h$^{-1}$Mpc. 

From Figure~\ref{fig:flux}, we see a clear difference in the trends of the observational results at low $q$. The earlier observations of A03 (open circles) show a definite increase in the transmitted flux, which is missing in the later results. In their paper, A05 note that the most likely reasons for this are the increase in sample size between A03 and A05, or an evolution in the LBGs between the epochs of the two data sets, taken at $z\sim 3$ and $z\sim 2$. Our simulation results (black line) show no proximity effect like that in A03, but agree closely with A05, indicating that sample size was the cause of the discrepancy, but not evolution since our data is at the earlier epoch of $z\sim 3$.

\begin{figure}
\centering
\includegraphics[width=7.0cm]{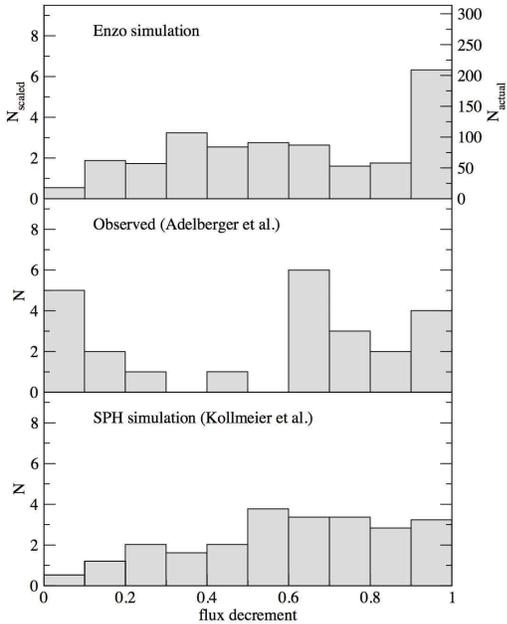}
\caption{Comparison between the simulated and observed Lyman-$\alpha$ absorption within 1\,Mpc of LBGs. The abscissa shows the flux decrement $1-<F_{1\rm Mpc}>$, where $<F_{1\rm Mpc}>$ is the average flux within 1\,Mpc of the galaxy. The left-hand ordinate shows the relative (in the case of simulated data) number of measurements at a given decrement interval. The right-hand ordinate on the top panel shows the actual number of measurement for the \emph{Enzo} simulation. The observational and SPH simulation results were taken from A05. }
\label{fig:flux_decrement}
\end{figure}

The distribution in absorbing gas can be further studied by looking at how it varies within 1\,h$^{-1}$Mpc of the LBGs. Figure~\ref{fig:flux_decrement} shows the distribution of the flux decrement, $1-\left<F\right>$, within this volume for the simulation results from \emph{Enzo}, the recent observations from A05 and the simulation results performed by Kollmeier et al., (taken from A05). Of the 2000 lines-of-sight that we passed through the simulation refined region, 884 passed within 1\,h$^{-1}$Mpc of an LBG. The average flux decrement within this area was calculated for each of these lines and scaled as for Figure~\ref{fig:flux}. A normalised scale, with the total number of measurements the same as Adelberger's observations, is marked on the left-hand axis while the actual numbers are shown on the right.

The observations indicate a bimodal distribution, with gas either very dense near to the galaxy, or largely absent. This, A05 argues, is consistent with the idea that anisotropic winds are present in these systems which are clearing a path in the gas which roughly half the lines-of-sight are passing through. If true, this could explain the discrepancy between the results, with the earlier 2003 observations picking out voids on one side of the galaxy that the larger sample set in 2005 averages out. 
The \emph{Enzo} simulation shows some indication of a bimodal distribution, but with very few transparent lines-of-sight.  In addition, there are a larger number of lines-of-sight with very high decrements. Kollmeier's results are similar, also seeing a lack of voids in the gas, but with no indication of a bimodal distribution at all. The simulations are therefore not seeing as strong a population of galaxies with the anisotropic voids.

\subsection{Intergalactic metals}

\begin{figure}
\begin{center}
\includegraphics[width=7.0cm]{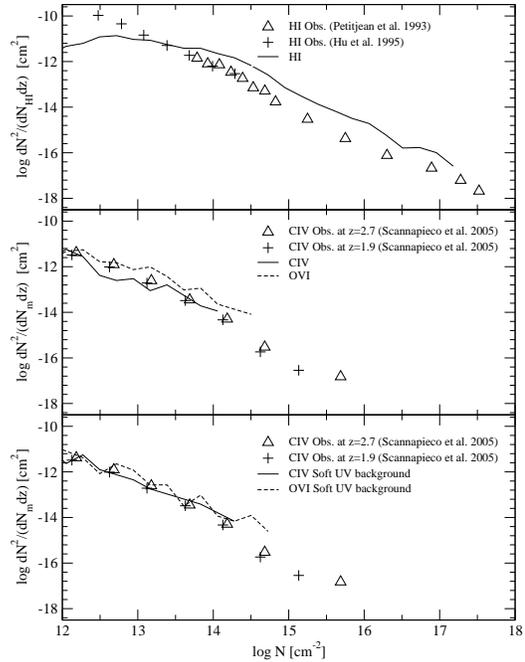}
\end{center}
\caption{Column density distributions for HI (top panel), CIV and OVI (middle panel) and for CIV and OVI calculated with a softer UV background along with observational data (as listed).}
\label{fig:metals}
\end{figure}

To measure the quantity of metals being ejected into the IGM, we calculate the column density of HI, CIV and OVI along 500 lines-of-sight.  These are plotted in Figure~\ref{fig:metals} along with observational data.  We note that we do not attempt to model observational effects such as confusion between HI and OVI lines, which will lead to systematic uncertainties at the low column-density end. 

The top panel of Figure~\ref{fig:metals} shows the column density of HI in our simulation (black line) alongside the observational results from \citet{Petitjean1993} and \citet{Hu1995}. Our results seem to be systematically slightly higher at medium to high column densities and curve away at lower values. The former is due to our choice of a high density area of the simulation box, where a number of LBG candidates were placed. The fall off at low densities is most likely due to a limitation in resolution of the simulation \citep{BryanMachacek1999} and low column density absorbers (which correspond to small filaments in under-dense regions) are underestimated. 

In the middle panel of Figure~\ref{fig:metals}, the column densities for CIV and OVI are plotted with the observational data for CIV from \citet{Scannapieco2005} at $z=2.7$ and $z=1.9$.
The match between our results and the observations at first seems extremely good, pointing towards the objects in the region producing the required metal enrichment of the IGM. However, as we have seen from the hydrogen distribution above, we have picked an over-dense region to study and therefore on average, we would expect the metal content to be lower. This implies that either the metals are more clustered around the galaxies than observed, or not enough metals are being ejected into the IGM.  

Softening the UV background, as shown in the bottom panel of Figure~\ref{fig:metals}, improves things slightly, but not enough to match the HI over-density.  Further softening, or a supersolar yield, could resolve the difference, but appear to be unlikely \citep{Aguirre2005}. Therefore, we conclude that either our galaxies do not have powerful enough outflows to drive the metals into the IGM, or they are not the main source of IGM enrichment. If they are not the source, it opens the question as to what is causing the metal enrichment. As can be seen from Figure~\ref{fig:projections}, contributions to our measured metal density come not just from the six LBG halos discussed in the previous section, but from all resolved galaxies. Therefore, if these are not enough, then smaller objects, below our minimum resolved halo mass of approximately $5\times 10^9$\,M$_\odot$, must be responsible. This minimum lies in the dwarf galaxy mass scale, so possible objects have to be dwarf-sized or smaller.

If this result is correct and LBGs are not the source of IGM enrichment, then the problem is left as to why there seems to be a correlation between LBGs and CIV systems (A03, A05). A solution to this was suggested by \citet{Porciani2005}, who propose early-forming dwarf galaxies, with the same biased distribution as LBGs, as the metal source. The LBGs are gravitationally drawn to the same over-dense sites and form in bubbles of old metals. \citet{Madau2001} also find that metal enrichment is more likely to come from pregalactic outflows than later ones from LBGs. Additionally, simulations performed by \citet{Scannapieco2005} tried multiple schemes of metal injection but found only when the metals were placed in bubbles round the galaxies were the observations replicated well. They point out that outflows from the smaller gravitational potentials from dwarves are a more likely candidate for the bubbles than the LBGs.\\*[0.05cm]

An alternative explanation for the LBGs weak effect on the IGM is that our feedback prescription is not able to correctly reproduce outflows of gas and metals.  It is clear that we cannot resolve the full multi-phase nature of the ISM, although we are currently working to understand and improve our treatment of feedback \citep{Tasker2006}. A visual inspection of our galaxies reveals several outflows, with the maximum velocity in the region being $750$\,km s$^{-1}$, encouragingly close to the observed outflows of  $775$\,km s$^{-1}$ by \citet{Pettini2002}. However, only two of our six halos showed strong outflows and only in one direction, indicating our outflows are neither homogeneous nor ubiquitous. Whether this is a problem is hard to determine; work performed by \citet{Kollmeier2005} and \citet{Theuns2002} on the effect of winds on absorption indicates that the lack of strong winds is probably not the cause of the decreased flux seen in the Figure~\ref{fig:flux}. \citet{Kollmeier2005} found that the gas in the IGM is unaffected by the presence of winds and has only a small impact on the optical depth close to LBGs. \citet{Theuns2002} and \citet{Bruscoli2003} likewise found that the winds preferentially expanded into the voids, leaving the hydrogen filaments intact. It therefore seems likely that we are correctly modeling the HI around the LBGs. 

The lack of metal production is more difficult to judge. From Figure~\ref{fig:projections}, the metals appear closely centered around their source, an effect that might be genuine due to pressure from the IGM \citep{Ferrara2005} or due to weak feedback. By using an extremely soft UV background and preventing heated gas from cooling for $10^7$\,yr, \citet{Theuns2002} is able to reproduce the observed CIV density well, implying that galactic outflows at $z=3$ can enrich the IGM if feedback is more efficient than presented here. However, \citet{Aguirre2005} find that the metals are too inhomogenously distributed compared to observations, a symptom that would not change in the presence of strong winds. They therefore argue that a second, high-z, enrichment mechanism is still needed.

\section{Conclusions}

We performed detailed simulations of galaxies formed at $z=3$ using an adaptive-mesh code which included recipes for star formation, metal production and SN feedback. From this data, we looked at the transmitted HI flux in simulated QSO spectra near large galaxies and the production of metals that are returned to the IGM.

Our calculated HI flux in the neighborhood of the galaxies agrees well with the most recent observational results (A05) in which the mean transmitted flux decreases near Lyman-break galaxies.  While the mean properties agree well, there is an indication that the distribution of transmitted fluxes does not show the same strong bimodality seen in the observations.

We also generate simulated CIV and OVI absorption lines and find that, in our simulation box, the column density distributions for these tracers are somewhat lower than observed.  A visual inspection of the simulations shows the metals to be quite concentrated around galaxies (both large and small). We conclude that while we do find a significant amount of the IGM polluted with metals, the amount falls short of observations, implying that either our treatment of feedback does not generate sufficiently strong winds, or that objects smaller than our minimum resolution of about $ 5\times 10^9$\,M$_\odot$ are responsible for the remaining IGM enrichment.

\acknowledgements

EJT and GLB acknowledge support from PPARC and the Leverhulme Trust.  This work was partially supported by NSF grant AST-0507161.  We thank the National Center for Supercomputing Applications for computing resources.


\begin{thebibliography}{}

\bibitem[Adelberger et al.(1998)]{Adelberger1998} Adelberger, K.~L., 
Steidel, C.~C., Giavalisco, M., Dickinson, M., Pettini, M., \& Kellogg, M.\ 
1998, \apj, 505, 18 
\bibitem[Adelberger et al.(2003)]{Adelberger2003} Adelberger, K.~L., 
Steidel, C.~C., Shapley, A.~E., \& Pettini, M.\ 2003, \apj, 584, 45 
\bibitem[Adelberger et al.(2005)]{Adelberger2005} Adelberger, K.~L., Shapley, A.~E., Steidel, C.~C., Pettini, M., Erb, D.~K., \& Reddy, N.~A.\ 
2005, \apj, 629, 636 
\bibitem[Aguirre et al.(2005)]{Aguirre2005} Aguirre, A., Schaye, 
J., Hernquist, L., Kay, S., Springel, V., \& Theuns, T.\ 2005, \apjl, 620, 
L13 
\bibitem[Bruscoli et al.(2003)]{Bruscoli2003} Bruscoli, M., Ferrara, 
A., Marri, S., Schneider, R., Maselli, A., Rollinde, E., \& Aracil, B.\ 
2003, \mnras, 343, L41
\bibitem[Bryan(1999)]{Bryan1999}Bryan,G.L. Comp. Phys. and Eng. 1999, 1:2, p.
\bibitem[Bryan et al.(1999)]{BryanMachacek1999} Bryan, G.~L., Machacek, 
M., Anninos, P., \& Norman, M.~L.\ 1999, \apj, 517, 13 
\bibitem[Croft et al.(2002)]{Croft2002} Croft, R.~A.~C., Hernquist, L., Springel, V., Westover, M., \& White, M.\ 2002, \apj, 580, 634 
\bibitem[Eisenstein \& Hut(1998)]{Eisenstein1998} Eisenstein, D.~J., 
\& Hut, P.\ 1998, \apj, 498, 137 
\bibitem[Ferland et al.(1998)]{Ferland1998} Ferland, G.~J., 
Korista, K.~T., Verner, D.~A., Ferguson, J.~W., Kingdon, J.~B., \& Verner, 
E.~M.\ 1998, \pasp, 110, 761 
\bibitem[Ferrara et al.(2005)]{Ferrara2005} Ferrara, A., 
Scannapieco, E., \& Bergeron, J.\ 2005, \apjl, 634, L37
\bibitem[Haardt \& Madau(2001)]{Haardt2001} Haardt, F., \& Madau, P.\ 2001, Clusters of Galaxies and the High Redshift Universe Observed in X-rays, 
\bibitem[Haardt \& Madau(1996)]{Haardt1996} Haardt, F., \& Madau, 
P.\ 1996, \apj, 461, 20 
\bibitem[Hu et al.(1995)]{Hu1995} Hu, E.~M., Kim, T.-S., Cowie, L.~L., Songaila, A., \& Rauch, M.\ 1995, \aj, 110, 1526 
\bibitem[Kollmeier et al.(2005)]{Kollmeier2005} Kollmeier, J.~A., Miralda-Escude, J., Cen, R., \& Ostriker, J.~P.\ 2005, ArXiv Astrophysics e-prints, arXiv:astro-ph/0503674 
\bibitem[Madau et al.(2001)]{Madau2001} Madau, P., Ferrara, A., 
\& Rees, M.~J.\ 2001, \apj, 555, 92 
\bibitem[O'Shea et al.(2004)]{OShea2004} O'Shea, B.~W., Bryan, G., Bordner, J., Norman, M.~L., Abel, T., Harkness, R., \& Kritsuk, A. 2004, ArXiv Astrophysics e-prints, astro-ph/0403044 
\bibitem[Petitjean et al.(1993)]{Petitjean1993} Petitjean, P., Webb, 
J.~K., Rauch, M., Carswell, R.~F., \& Lanzetta, K.\ 1993, \mnras, 262, 499
\bibitem[Pettini et al.(2001)]{Pettini2001} Pettini, M., Shapley, A.~E., Steidel, C.~C., Cuby, J.-G., Dickinson, M., Moorwood, A.~F.~M., Adelberger, K.~L., \& Giavalisco, M.\ 2001, \apj, 554, 981 
\bibitem[Pettini et al.(2002)]{Pettini2002} Pettini, M., Rix, S.~A., Steidel, C.~C., Adelberger, K.~L., Hunt, M.~P., \& Shapley, A.~E.\ 2002, \apj, 569, 742
\bibitem[Porciani \& Madau(2005)]{Porciani2005} Porciani, C., \& Madau, P.\ 2005, \apjl, 625, L43
\bibitem[Scannapieco et al.(2005)]{Scannapieco2005} Scannapieco, E., 
Pichon, C., Aracil, B., Petitjean, P., Thacker, R.~J., Pogosyan, D., 
Bergeron, J., \& Couchman, H.~M.~P.\ 2005, ArXiv Astrophysics e-prints, 
arXiv:astro-ph/0503001 
\bibitem[Schaye et al.(2003)]{Schaye2003} Schaye, J., Aguirre, A., Kim, T.-S., Theuns, T., Rauch, M., \& Sargent, W.~L.~W.,\ 2003, \apj, 596, 768 
\bibitem[Songaila(2001)]{Songaila2001} Songaila, A.\ 2001, \apjl, 561, L153 
\bibitem[Tasker \& Bryan(2006)]{Tasker2006} Tasker, E. \& Bryan, G.L. 2006, \apj, in press
\bibitem[Tassis et al.(2003)]{Tassis2003} Tassis, K., Abel, T., Bryan, G.~L., \& Norman, M.~L., 2003, \apj, 587, 13 
\bibitem[Theuns et al.(2002)]{Theuns2002} Theuns, T., Viel, M., 
Kay, S., Schaye, J., Carswell, R.~F., \& Tzanavaris, P.\ 2002, \apjl, 578, 
L5 

\end{thebibliography}
\end{document}